\documentclass{kluwer}    

\newdisplay{guess}{Conjecture}

\input psfig.sty

\def\la{\mathrel{\mathchoice {\vcenter{\offinterlineskip\halign{\hfil
$\displaystyle##$\hfil\cr<\cr\sim\cr}}}
{\vcenter{\offinterlineskip\halign{\hfil$\textstyle##$\hfil\cr
<\cr\sim\cr}}}
{\vcenter{\offinterlineskip\halign{\hfil$\scriptstyle##$\hfil\cr
<\cr\sim\cr}}}
{\vcenter{\offinterlineskip\halign{\hfil$\scriptscriptstyle##$\hfil\cr
<\cr\sim\cr}}}}}

\def\ucm{\mathrel{\mathchoice {\vcenter{\offinterlineskip\halign{\hfil
$\displaystyle##$\hfil\cr\mbox{\small\sc uc}\cr\mbox{\small\sc m}\cr}}}
{\vcenter{\offinterlineskip\halign{\hfil$\textstyle##$\hfil\cr
\mbox{\small\sc uc}\cr\mbox{\small\sc m}\cr}}}
{\vcenter{\offinterlineskip\halign{\hfil$\scriptstyle##$\hfil\cr
\mbox{\small\sc uc}\cr\mbox{\small\sc m}\cr}}}
{\vcenter{\offinterlineskip\halign{\hfil$\scriptscriptstyle##$\hfil\cr
\mbox{\small\sc uc}\cr\mbox{\small\sc m}\cr}}}}}

\def\reduceme{R$_{\mbox{\small\rm E}}$D$\ucm$E}

\begin{document}                                          

\begin{article}
\begin{opening}         
\title{Line Strengths and Line Strength Gradients in Bulges along
	the Hubble Sequence\thanks{Based on observations obtained at ESO, La
	Silla, Chile (Observing Programmes 58.A-0192, 	59.A-0774, and
	61.A-0326), and at the Isaac Newton Telescope operated 
	on the island of La Palma by the Isaac Newton Group in the Spanish
	Observatorio del Roque de los Muchachos of the Instituto de
	Astrofisica de Canarias}}  
\author{Paul \surname{Goudfrooij}}
\institute{Space Telescope Science Institute, 
	3700 San Martin Drive, 
	Baltimore, MD 21218, United States of America
	\email{goudfroo@stsci.edu} 
	\\ [0.3ex] 
	Affiliated with the Astrophysics Division, SSD, 
	European Space Agency} 
\author{Javier \surname{Gorgas}}
\institute{Departamento de Astrofisica, Facultad de Ciencias Fisicas,
        Universidad Complutense de Madrid, E-28040 Madrid, Spain
        \email{fjg@astrax.fis.ucm.es}}
\author{Pascale \surname{Jablonka}}
\institute{DAEC--URA173, Observatoire de Paris-Meudon, Place Jules
	Janssen, Meudon, F-92195, France \email{Pascale.Jablonka@obspm.fr}}
\runningauthor{P.\ Goudfrooij, J.\ Gorgas, \& P.\ Jablonka}
\runningtitle{Stellar Populations within Bulges along the Hubble
Sequence} 

\begin{abstract} 
We present first results of a comprehensive survey of deep long-slit
spectra along the minor axis of bulges of edge-on spiral galaxies. 
Our results indicate that stellar populations in bulges  are fairly
old and encompass a range of metallicities. The luminosity-weighted
ages of bulges range from those found for cluster ellipticals to 
slightly ``younger'' (by up to only a few Gyr, however). Their  
$\alpha$/Fe element ratio is typically supersolar, consistent with
those found in giant ellipticals. The radial line-strength gradients
in bulges correlate with bulge luminosity. Generally, these findings
are more compatible with predictions of the ``dissipative collapse'' model
than with those of the ``secular evolution'' model. 
\end{abstract}

\end{opening}           
\vspace*{-4.ex}
\section{Introduction: The Formation of Bulges of Spirals}  

\vspace*{-1.2ex} 
Bulges of spiral galaxies are cornerstones for constraining
theories of galaxy formation. 
Located at the centers of spiral galaxies, they hold the signature of the
sequence of formation ---outside-in or inside-out--- of the
different sub-systems building a spiral galaxy:\ halo, disc, and bulge. The
prominence of bulges varies widely along the Hubble sequence. This contrasts
with the situation for spiral disks whose mass is nearly constant among
all types of spirals \cite{arijab92}.  Hence, bulges constitute a main key to
our understanding of spiral galaxy evolution. 
To set the context for our project, we describe below the two currently most
popular scenarios on bulge formation that have been proposed over the
years. \smallskip 

\noindent
{\bf Monolithic Dissipative Collapse:} \,Bulges form before disks do, 
on very short time 
scales. As the pregalactic gas collapses and forms stars, metals are ejected
by winds blown by massive stars and supernovae. 
If the galaxy potential well is deep
enough to retain these ejecta, the enriched gas is carried inward
to the galaxy center. As new stars are formed, their chemical
composition reflects that of the gas. The result is a radial
metallicity gradient, whose amplitude should increase with galaxy mass and
luminosity 
(Carlberg, \citeyear{carl84}; Arimoto \& Yoshii, \citeyear{ariyos87}). 

Evidence in favor of this scenario has been
presented by means of spectroscopic studies of {\it ellipticals}:
\inlinecite{caro+93} reported the presence of a 
correlation between radial Mg$_2$ absorption-line-strength
gradients and galaxy mass, velocity dispersion, and central line
strength for low-- to intermediate-mass ({\it i.e.}, $\cal{M} \la
\mbox{10$^{11}$} 
\cal{M}_{\odot}$) ellipticals, as opposed to {\it giant\/} ellipticals (but
see different results by Gonz\'alez \& Gorgas \citeyear{gongor96} and
Kobayashi \& Arimoto 
\citeyear{kobari99}). The gradients may thus be driven mostly by 
dissipation in smallish ellipticals, whereas merging and violent relaxation
play a more important role for giant ellipticals.  
What about the actual case of bulges of spirals\,? There is strong
evidence that bulges are analogous to low-- to
intermediate-mass ellipticals in several {\it global\/}
principal properties: they populate the same 
location in the Fundamental Plane \cite{bend+92}; they form a continuous
sequence in the $V_{max}/\sigma_0$ vs.\ ellipticity diagram (being
supported by rotation, Bender et al.\ 
\citeyear{bend+92}).  As to the stellar populations, \inlinecite{jabl+96}
performed a spectroscopic pilot study of the centers 
of bulges of face-on spirals.  They revealed the existence
of striking similarities between bulges and ellipticals. {\it E.g.}, their
luminosity-metallicity relations, when derived from
$\alpha$-elements (such as the Mg$_2$ index), are consistent with one
another. 

\smallskip

\noindent
{\bf Secular Evolution:\ } Bulges form from disc material through
 redistribution of angular momentum. In this scenario, large amounts of gas
 are driven into the central region of the galaxy by a stellar bar and
 trigger intense star formation ({\it e.g.}, Pfenniger \& Norman
 \citeyear{pfenor90}).
 If enough mass is  accreted, the bar 
 itself will be dissolved and the resulting galaxy will reveal a bigger bulge
 than before bar formation; galaxies would thus evolve from late to earlier
 types along the Hubble sequence.

This model also certainly has its attractions. Bars
appear in at least $\sim$\,2/3 of disk galaxies \cite{selwil93}, and several
numerical simulations have indicated that bars can significantly influence the
dynamical evolution of galaxies, through mechanisms such as disk thickening by
box-peanut or bending instabilities (Combes et al., \citeyear{comb+90})
or radial mass inflow towards the center \cite{friben95}. 
Since bars appear to be able to affect the global dynamics of galaxies, it is
natural to suspect that bars can be responsible for significant chemical
evolution as well, due to mixing by the bar-induced kinematics. 
Recent N-body simulations (that include the effect of star formation) have
shown this to be a valid suspicion in the sense that any initial abundance
gradient is indeed significantly washed out $\sim$\,1 Gyr after formation of a
bar, both for gas and stars (\opencite{frie+94}; \opencite{frie98}). The slope
of the abundance gradient is found to flatten beyond the co-rotation radius,
by up to $\sim$\,50\%\ through the effect of one (strong) bar. Observational
evidence for this effect exists, albeit  
only for the gas component so far. Several groups have shown that global
radial gradients of the gas metallicity (in terms of [O/H] in H\,{\sc ii}
regions) in barred spirals are shallower than gradients in ``normal'' spirals
of the same Hubble type (Vila-Costas \& Edmunds, \citeyear{viledm92}; Zaritsky
et al., \citeyear{zari+94}; Martin \& Roy, \citeyear{marroy94}). Furthermore, 
a clear  relation seems to exist between the relative length of the bar with
respect to the size of the (optical) disk and the slope of the radial [O/H]
gradient among barred spirals \cite{marroy94} in the sense that the larger the
bar/disk ratio, the shallower the abundance gradient. \medskip \smallskip 

\noindent
\centerline{\sc 1.1. Line-Strength Gradients: Key to the
Controversy\,?}  

\smallskip
\noindent
It thus seems that radial population gradients in bulges have great 
potential in discriminating between the main bulge formation scenarios. 
Previous work on stellar populations
in bulges has mainly been limited to photometric studies: Balcells \& Peletier
(\citeyear{balpel94}) in UBRI, and Terndrup et al.\ \shortcite{tern+94} in J 
and K, show that 
for luminous bulges, color gradients become increasingly negative with
increasing luminosity (similar to the case among ellipticals).
However, many faint bulges deviate from this trend and keep
showing strong negative color gradients, and the two studies mentioned
above do not converge in their conclusions. Besides, it is now well known
that broad-band photometry is incapable of accurately disentangling the
effects of age, metallicity, and dust absorption, whereas stellar line
strenghts are dust-independent and {\it can\/} separate the effects of age
and metallicity ({\it e.g.}, Worthey, \citeyear{wort94}, hereafter
W94; Vazdekis et al.,  
\citeyear{vazd+96}). 
With this in mind, we embarked on a spectroscopic survey of a significant
sample of bulges to compare their possible star formation histories with those
in ellipticals. \vspace*{-2.5ex}

\section{Description of our Project}

\vspace*{-1ex} 
In order to avoid contamination from disk light, we selected a sample of 28
edge-on, nearby spiral galaxies from the UGC and
ESO-LV catalogs, and oriented  
the spectrograph slit along the minor axis of the bulges. The sample
encompasses a considerable range in luminosities for each Hubble type 
($18.4 < -M_V < 21.5$). 
The observations (spectroscopy and two-color imaging) of the
northern spirals were obtained with the 2.5-m Isaac Newton Telescope at La
Palma, 
while the southern spirals were observed with the ESO NTT and 3.6-m
telescopes.  
As to the spectroscopy, the typical exposure time was $\sim$\,4 hours per
galaxy, and the instrumental resolution was typically of order 100 km
s$^{-1}$. The spectral range used (typically 3900\,--\,5500\AA)
allowed the measurement of most Lick/IDS line-strength indices. 
 
All spectral data reduction and line index measurements were performed using
the R$_{\mbox{\small\rm E}}$D$\ucm$E package\footnote{Cardiel \& Gorgas,
{\sf http://www.ucm.es/info/Astrof/reduceme/reduceme.html}} which
propagates errors associated with all reduction steps along with
the science data. The line indices were fully calibrated to the Lick/IDS
system (including broadening corrections), using measurements of $\sim$\,40
IDS standard stars \cite{gorg+93}. Full details of the sample selection, the
observations, and the calibrations will be provided in a
forthcoming paper.   \vspace*{-4ex}

\section{Results}

\vspace*{-1ex}
At the time this paper is written, data analysis of 16 bulges
in our sample is completed. Here we describe some highlights of the results
from this subsample. \medskip 

\noindent
\centerline{\sc 3.1.\ Central Line Strengths}

\medskip\noindent
In order to effectively compare ``central'' line strengths for bulges with
those of ellipticals in the literature, we extracted spectra with spatial
extent 4$''$ from the galaxy center {\it but avoiding the innermost dust
lane\/} to eliminate disk light.  
In disentangling the effects of age and metallicity, most authors
have used a combination of H$\beta$ and ``metallic'' line-strengths such as 
Mg$_2$, Mg\,$b$, Fe5270 and Fe5335 ({\it e.g.}, Gonz\'{a}lez
1993). However, the use of the H$\beta$ and Mg\,$b$ indices is very
limited in case nebular emission 
is present \cite{gouems96}, which is ``unfortunately'' often the case in our
spectra. We did not attempt to correct H$\beta$ for  emission, but rather used
the H$\gamma_A$ index as age-sensitive index (calibrated in age/metallicity
space by Worthey \& Ottaviani 1997). Since the relative strength of Balmer
line emission decreases rapidly with Balmer order ({\it e.g.},
Osterbrock 1974), H$\gamma_A$ is significantly less diluted by nebular
emission than H$\beta$ is.  

\begin{figure}
\centerline{
\psfig{file=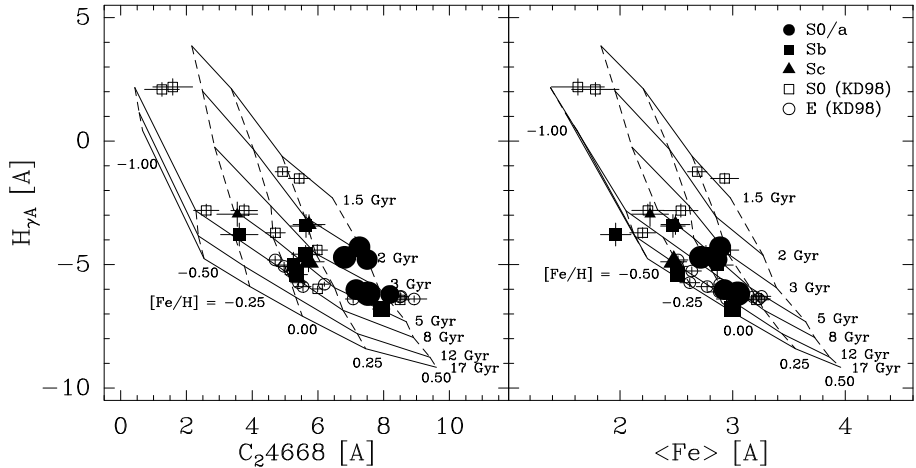,width=12cm,angle=-90.}
}
\caption[]{The age-sensitive index H$\gamma_{{\rm A}}$ is plotted against
metallicity indicators C$_2$4668 {\sl (left)\/} and $<\!\mbox{Fe}\!>$ {\sl
(right)} for the central regions of bulges in our 
sample. Age-metallicity grids from population synthesis models by Worthey
(1994) and Worthey \& Ottaviani (1997) are overplotted. Solid lines represent
constant age, while dashed lines represent constant metallicity. Symbol
definitions are shown in the inset; symbol size is proportional to the bulge
luminosity. Open circles and open squares represent centers of Fornax
ellipticals and S0s, respectively, from KD98.}
\label{f:metals_hga}
\end{figure}

In Fig.\ \ref{f:metals_hga} we present plots of the  H$\gamma_A$ index vs.\
two metallicity indicators:\ {\it (i)\/} C$_2$4668 which Jones \& Worthey
(\citeyear{jonwor95}) 
identified as a particularly metallicity-sensitive index, and {\it
(ii)\/} 
$<\!\mbox{Fe}\!>$ = (Fe5270+Fe5335)/2. For comparison, we also plot the 
results of Kuntschner \& Davies (\citeyear{kundav98}; hereafter KD98) for the
centers of E and S0 
galaxies in the Fornax cluster, as well as predictions from single-burst
population synthesis models (W94; Worthey \& Ottaviani 1997). It turns out
that the bulges in our sample have ages similar
to (or up to a few Gyr ``younger''\footnote{Recall however that
line-strength indices reflect {\it luminosity-weighted\/} properties in a
galaxy. A young population that is small in mass ---but relatively large in
luminosity---, can dramatically change the index values.} than) those of 
{\it cluster\/} ellipticals. On the other hand, luminosity-weighted ages of
a sample biased towards {\it field\/} ellipticals (Gonz\'alez 1993) span the
whole range found for 
bulges. The 
metallicities of bulges cover a range similar to those of 
ellipticals.   
Interestingly,  
bulges of later-type (Sb\,--\,Sc) spirals are, in the mean, less metal rich
than their counterparts in earlier-types. This seems to be a bulge
luminosity effect, judging from the symbol sizes in Fig.\
\ref{f:metals_hga}. There are no other obvious distinctions between bulges of
different Hubble types in this context.   

\begin{figure}
\centerline{
\psfig{file=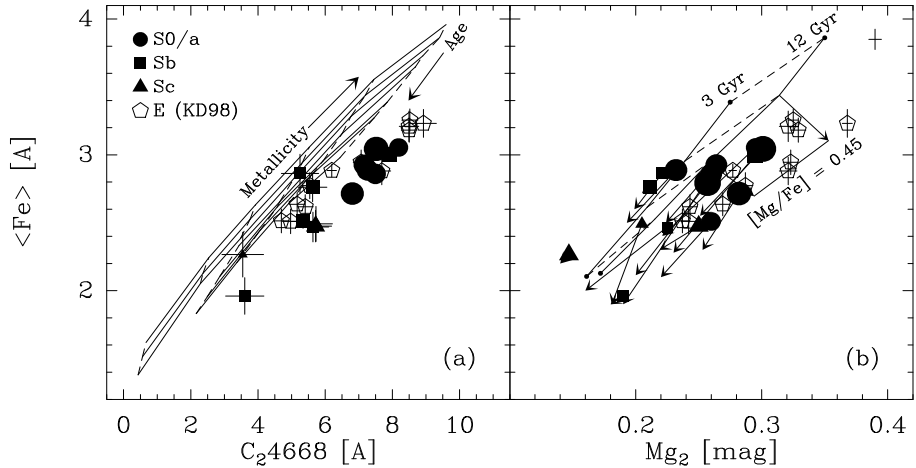,width=12cm,angle=-90.}
}
\caption[]{{\it (a)\/} C$_2$4668 vs.\ $<\!\mbox{Fe}\!>$ indices for the
central regions of bulges in our sample. Models by Worthey (1994) are
overplotted as in Fig.\ \ref{f:metals_hga}. Symbols are 
shown in the inset. The symbol size is proportional to the bulge
luminosity of the galaxies. 
Open pentagons represent Fornax ellipticals from Kuntschner \& Davies (1998).
{\it (b)\/} Mg$_2$ vs.\ $<\!\mbox{Fe}\!>$ indices for bulges
in our sample. Symbols as in Fig.\ \ref{f:alpha_fe}a. The filled symbols
represent the ``central'' indices of the bulges, and the arrows
pointing downwards from those symbols denote the radial (vertical) gradients
going outwards (length of arrow is proportional to the gradient
slope). Overplotted are models by Worthey (1994) [only for ages 3
and 12 Gyr; solid lines] and a correction for [Mg/Fe] = 0.45 for the 12 Gyr
isoage line (calculated from models by Weiss et al.\ 1995). }
\label{f:alpha_fe}
\end{figure}

Another interesting point is that, comparing the positions of the galaxies in
Figs.\ \ref{f:metals_hga}a and \ref{f:metals_hga}b, it appears that C$_2$
({\it i.e.}, the C$_2$4668 index) 
is overabundant in ellipticals and bulges with respect to solar abundance
ratios. This is illustrated in Fig.\ \ref{f:alpha_fe}a which compares
C$_2$4668 with $<\!\mbox{Fe}\!>$. Comparing the observations with the
superimposed models of W94, it is obvious that C$_2$4668 is stronger than
indicated by the models (which employed solar abundance ratios). Whether this
is a real overabundance effect (or, {\it e.g.}, due to a problem in the
fitting functions) is an issue which deserves further analysis. 
\newpage

\noindent
\centerline{\sc 3.2.\ Radial Line-strength Gradients} 

\begin{figure}
\centerline{
\psfig{file=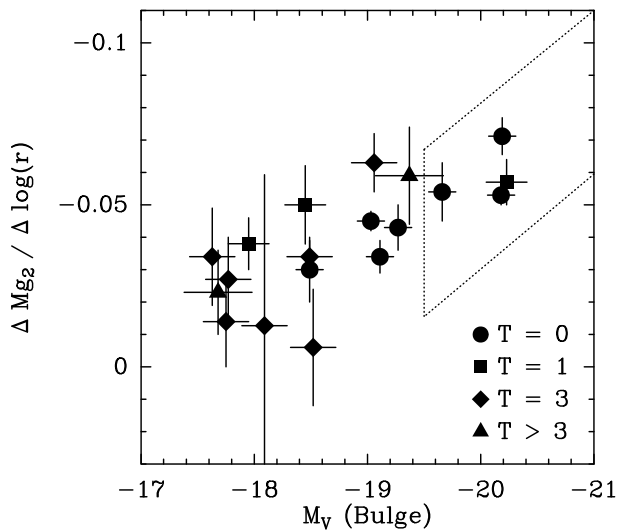,width=8cm}
}
\caption[]{Logarithmic radial gradient of Mg$_2$ index vs.\ absolute
$V$-band {\bf bulge} luminosity of the bulges analyzed to date. Symbols are
shown in the inset (where $T$ is the RC3 morphological type). The dotted
lines depict the extremes of the $\Delta$Mg$_2$/$\Delta$log($r$) vs.\ $M_V$ 
relation among low-- and intermediate-luminosity E and S0 galaxies from
Carollo et al.\ (1993).}
\label{f:Mv_vs_grad}
\end{figure}

\medskip\noindent
Fig.\ \ref{f:alpha_fe}b shows a $<\!\mbox{Fe}\!>$ vs.\ Mg$_2$
plot for the bulges in our sample. The symbols depict the ``central''
values, while the arrows point towards the outermost well-measured 
values; the length of the arrows is a measure of the slope of the
gradient. 
It is clear that, in most bulges, Mg is 
overabundant with respect to solar abundance ratios. The [Mg/Fe] abundance
ratio in bulges is similar to those found in ellipticals, and stays
more or less constant throughout the radial extent of bulges. From the
(overplotted) $\alpha$-element-overabundance models by \inlinecite{weis+95}
for [Mg/Fe] = 0.45 and an age of 12 Gyr (and  mixing length parameter
$\alpha_{\mbox{\footnotesize\sc mlt}}$\,=\,1.5), we estimate that [Mg/Fe]
$\la$ 0.4. \\ 
\hspace*{\parindent} Is there a correlation between the metallicity gradient
and luminosity for 
bulges, as predicted by dissipative collapse models\,? This
relation is depicted in Fig.\ \ref{f:Mv_vs_grad}. Bulge
luminosities\footnote{using $H_0$ = 50 km s$^{-1}$ Mpc$^{-1}$} were derived
by performing ellipse fits to the isophotes of the galaxy 
images, after masking out wedges encompassing the dusty disks ($\pm$
20$^{\circ}$ from the major axes). The gradient-luminosity correlation
indeed exists; any further distinction 
between bulges of different Hubble types within this relation is not
obvious. Incidentally, the Mg$_2$ gradients of the most luminous bulges as
well as the slope of the gradient-luminosity relation are similar to 
those among low-luminosity ellipticals in the Carollo et al.\ (1993) sample.
\vspace*{-3ex}

\section{Concluding Remarks}

\vspace*{-1ex}
From the spectral data of bulges of spirals in our sample analyzed so far, we
have established the following main results: \\ [-4ex]
\begin{enumerate}
\item 
Bulges have luminosity-weighted metallicities varying from roughly
$-0.50$ to $+0.20$ in [Fe/H], as measured from H$\gamma_A$ vs.\ $<$\,Fe\,$>$
index diagrams. Many bulges are as old as cluster ellipticals, but some
(low-luminosity) bulges have luminosity-weighted ages up to a few Gyr younger
(note that this result may be influenced by residual light from thick 
disks).  
 \\ [-4.2ex]
\item 
Bulges are typically overabundant in $\alpha$-elements, up to
[Mg/Fe] $\simeq$ +0.4 dex, throughout their radial extent. There is no
obvious correlation between [Mg/Fe] and bulge luminosity. This
is similar to the situation among ellipticals, and indicates that the
bulk of the stars in bulges typically formed within a few Gyr 
(before the onset of SNIa explosions; {\it e.g.}, Worthey, Faber \&
 Gonz\'alez \citeyear{wort+92}).   
 \\ [-4.2ex]
\item 
There is a correlation between the radial metal-line index gradients and the
bulge luminosities for bulges in our sample. \\ [-4ex]
\end{enumerate}
These first results seem to be generally more compatible with the predictions
of the ``dissipative collapse'' models than with those of the ``secular
evolution'' models (cf.\ Sect.\ 1). However, some individual
bulges seem to be younger than the rest and show shallow radial Mg$_2$
gradients, which in turn can be due to effects induced by bar instabilities. 
Moreover, we postpone the announcement of ``final'' conclusions until the
data of our full sample of 28 bulges has been analyzed.

\medskip\noindent
{\bf Acknowledgements.}~
We thank the referee, Richard de Grijs, for his timely and critical review of
this paper. We are grateful to Nicol\'{a}s Cardiel for his help with the
\reduceme\ package, and to Harold Kuntsch\-ner for giving access to his data
in electronic form. \vspace*{-3.ex}

%

\end{article}
\end{document}